\documentclass[article]{aa}

\renewcommand\makeLineNumber{}

\usepackage{graphicx}
\usepackage{txfonts}
\usepackage{lipsum}
\usepackage{subcaption}         
\usepackage{lscape}             
\usepackage{placeins}           

\usepackage[dvipsnames]{xcolor}
\usepackage{hyperref}
\hypersetup{pdfborder = {0 0 0},
    colorlinks = true,
    linkbordercolor = {white},
    citecolor=NavyBlue,
    urlcolor=Green
}  

\newcommand{\orcidlink}[1]{\protect\href{https://orcid.org/#1}{\protect\includegraphics[width=8pt]{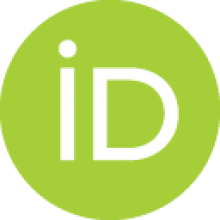}}}

\usepackage{xspace}
\newcommand{\kp}{\emph{Kepler}\xspace}
\newcommand{\ie}{i.e.\@\xspace} 
\newcommand{\eg}{e.g.\@\xspace} 
\renewcommand{\eqref}[1]{Eq.~\ref{#1}}
\newcommand{\fref}[1]{Fig.~\ref{#1}}
\newcommand{\Fref}[1]{Figure~\ref{#1}}
\newcommand{\teff}{\ensuremath{T_{\rm eff}}\xspace}

\newcommand{\logg}{\ensuremath{\log g}\xspace}

\usepackage{natbib,twoopt}
\bibpunct{(}{)}{;}{a}{}{,} 

\begin{document}

   \title{Optimising the global detection of solar-like oscillations}
   \subtitle{Tuning the frequency range for asteroseismic detection predictions and searches}    \titlerunning{Optimising the global detection of solar-like oscillations}

%

    \author{Mikkel N. Lund\inst{1}\fnmsep\thanks{mikkelnl@phys.au.dk}\orcidlink{0000-0001-9214-5642}
        \and William J. Chaplin\inst{2}\fnmsep\thanks{w.j.chaplin@bham.ac.uk}\orcidlink{0000-0002-5714-8618}
        }

   \institute{Stellar Astrophysics Centre, Department of Physics and Astronomy, Aarhus University, Ny Munkegade 120, DK-8000 Aarhus C, Denmark\\
    \and
              School of Physics and Astronomy, University of Birmingham, Birmingham, B15 2TT, United Kingdom\\
    }
  
   \authorrunning{Lund \& Chaplin}
   \date{Received 20 January 2026, Accepted 3 March 2026}

  \abstract
   {A well-established method exists for predicting the detectability of solar-like oscillations that has seen extensive application supporting target selection strategies for space-based photometric missions. The method assesses the probability of making an asteroseismic detection based on the expected global signal-to-noise ratio (SNR) of the observed signal due to the oscillations against the broadband background due to shot noise and granulation. Known stellar parameters are used to compute the expected oscillation and granulation signal, while instrumental specifications and the apparent brightness of the target are used to compute the expected shot noise.}
   {We explore whether there is an optimal choice for the range in frequency, $W$, over which the global SNR is determined. The observed power in solar-like oscillations is assumed to follow a Gaussian-like envelope of the full width at half maximum $\Gamma_{\rm env}$, centred on the frequency of maximum oscillation power. It has been common practice to set $W \simeq 2\Gamma_{\rm env}$ when predicting detections.} 
   {We make numerical predictions of the global SNR and resulting detection probabilities for a range of underlying stellar and observational parameters, adopting different choices for the width using $W =\alpha \Gamma_{\rm env}$, where $\alpha$ is a multiplicative coefficient that controls the width. We also explored the effect of this choice on detection yields across an ensemble of targets, using a sample of bright solar-like oscillators observed by TESS as a representative example.}
   {We found that the commonly adopted value of $\alpha \simeq 2$ is a sub-optimal choice and that adopting a range with $\alpha \simeq 1.2$ maximises the detection probability. There can also be a substantial effect on the predicted detection yield across a sample of stars.}
   {In summary, we recommend the adoption of a range $W \simeq 1.2\Gamma_{\rm env}$, not only in computations of the detection probabilities, but also in actual searches for oscillations in real data based on testing the significance of excess mode power, since its adoption will also optimise the probability of making robust detections.}


   \maketitle

\section{Introduction}
\label{sec:intro}

Solar-like oscillations have now been detected in hundreds of cool main-sequence and subgiant stars and in thousands of red giants (\eg see \citealt{Chaplin2013, Garcia2019}). Because the oscillation parameters follow scaling relations that depend on fundamental stellar properties to a very good approximation (\eg see \citealt{BasuChaplin2017}), it is possible to make straightforward predictions of the detectability of the oscillations if the noise budget of the observations is known and to use these known characteristics to tune codes that seek to detect signatures of oscillations in the data.  The ability to make such predictions is central to target selection strategies, in particular those for space-based photometric missions. It can inform decisions on the selection of observing fields, specific targets, and observing durations, and so on in order to fulfil mission-specific requirements on detection yields that support specific science goals.

We present an optimisation of the detection prediction method of \citet{Chaplin2011}. This approach was developed to inform asteroseismic target selection strategies for the NASA \kp mission, and has since been used for the K2 \citep{Chaplin2015,Lund2016,Lund2024}, the Transiting Exoplanet Survey Satellite (TESS) \citep{Campante2016,Schofield2019,Hey2024}, and the CHaracterising ExOPlanet Satellite (CHEOPS) \citep{Moya2018} missions, and more recently, in preparations for the upcoming ESA PLAnetary Transits and Oscillations of stars (PLATO) mission \citep{Goupil2024}. The optimisation is of the choice of frequency range over which to calculate the global signal-to-noise ratio (SNR) of the oscillations in the frequency power spectrum of the data. This optimal choice of range is then also relevant to the actual searches for oscillations in real data, since its adoption optimises the probability of making robust detections. 

The layout of the rest of the paper is as follows. In Sect.~\ref{sec:recipe} we recapitulate the key elements of the method. Sect.~\ref{sec:range} contains the main results of the paper. Sect.~\ref{sec:forms} presents formulae to describe the integrated oscillations power and global SNR over different ranges in frequency. We then use these formulae in Sect.~\ref{sec:opt} to show that the commonly adopted choice of frequency range is sub-optimal, and we provide updated guidance on the best range to adopt in detection predictions and in actual searches for the oscillations. We present our main conclusions in Sect.~\ref{sec:conc}.

\section{The detection recipe}
\label{sec:recipe}

We begin by summarising the key elements of the \citet{Chaplin2011} approach. First, we followed the well-established convention of assuming that when smoothed in frequency, the observed power in solar-like oscillations follows a Gaussian-like envelope centred on the frequency of maximum oscillation power, $\nu_{\rm max}$. We may then write the power spectral density $P(\nu)$ of the envelope as
 \begin{equation}
  P(\nu) = H_{\rm env} \exp\!\left[ -\,4\ln 2 \left( \frac{\nu - \nu_{\max}}{\Gamma_{\rm env}} \right)^2 \right],
 \end{equation}
where $H_{\rm env}$ and $\Gamma_{\rm env}$ are the height and full width at half maximum (FWHM) of the envelope, respectively. The total area below the envelope corresponds to the total detected mean-square power in the oscillations and is given by \citep[\eg see][]{BasuChaplin2017}
 \begin{equation}
 P_{\rm tot} = \left( \frac{\pi}{4\ln 2} \right)^{1/2} H_{\rm env} \Gamma_{\rm env}\, .
 \label{eq:ptot}
 \end{equation}
The height of the envelope is, in turn, given by
  \begin{equation}
  H_{\rm env} = \frac{\varsigma A_{\rm max}^2 
  }{\Delta\nu}\, ,
 \end{equation}
where $A_{\rm max}$ is the \textsc{rms} maximum equivalent radial mode amplitude, $\Delta\nu$ is the large frequency separation, and $\varsigma$ is the sum of the mode visibilities in power, that is, the sum over $l=0$, 1, 2, and 3 in each order, normalised to unity for $l=0$. The width of the envelope was assumed to take the form (\eg see \citealt{Mosser2010, Mosser2012, Chaplin2011})
  \begin{equation}
  \Gamma_{\rm env} = \alpha \nu_{\rm max}^{\beta}\, ,
 \end{equation}
where $\alpha$ and $\beta$ are coefficients. Combining the above gives
 \begin{equation}
 P_{\rm tot} = \left( \frac{\pi}{4\ln 2} \right)^{1/2} \alpha \varsigma\,A_{\rm max}^2 \frac{\nu_{\rm max}^{\beta}}{\Delta\nu}\, .
 \label{eq:ptot}
 \end{equation}
Predictions of $P_{\rm tot}$ may be made for a given stellar target using well-established scaling relations for the oscillation parameters (e.g. see \citealt{BasuChaplin2017} for further details). 

The \citet{Chaplin2011} detection method then uses the global SNR in the oscillations, as defined by
 \begin{equation}
 {\rm SNR_{\rm tot}} = \frac{P_{\rm tot}}{B_{\rm tot}}\, ,
 \end{equation}
where $B_{\rm tot}$ is the total background power across the frequency range occupied by the modes. In what follows, we denote this range as $W$. 

The background has contributions from shot and other instrumental noise, as well as from stellar granulation. We assumed that the underlying limit spectra arising from the combination of background sources, that is, the spectra that would be given by an infinite number of realisations of the data, vary slowly or are approximately constant across the frequency range of interest. The larger the shot-noise contribution, the more the background power spectral density will tend to be flat in frequency, because this white contribution will then dominate the frequency-dependent granulation term. In practice, ignoring the frequency dependence of the background does not significantly affect the detection probability calculation (\citealt{Chaplin2011}).

We then specified an average background power spectral density $\bar{b}$, from which it follows that
 \begin{equation}
 B_{\rm tot} = \bar{b} W\, .    
 \end{equation}
Estimates of the noise will depend on the apparent brightness of the target, the instrumental response, and the sensitivity; while the granulation contribution is computed at $\nu_{\rm max}$ using scaling relations that depend on fundamental stellar properties \citep{Kallinger2014,Larsen2026}.

When the time-domain observations span a duration $T$, the number of independent frequency bins $N$ in the range $W$ is
 \begin{equation}
 N = W T.
 \label{eq:N}
 \end{equation}
The power spectral density within any independent bin will follow $\chi^2$ 2 d.o.f. statistics about the underlying limit spectrum \citep{Appourchaux2014}. The global SNR, which is computed over $N$ bins, then follows $\chi^2$ $2N$ d.o.f. statistics. With predicted estimates of the underlying $P_{\rm tot}$ and total noise $B_{\rm tot}$ in hand for a particular target, we may then calculate the probability $p_{\rm final}$ that the observed SNR would exceed some false-alarm threshold $p_{\rm false}$, consistent with the statistics defined above \citep{Appourchaux2004}.

Whilst the various implementations of the \citet{Chaplin2011} detection procedure (\ie in \citealt{Campante2016, Moya2018, Schofield2019, Hey2024, Goupil2024}) have made varying assumptions about how $\Gamma_{\rm env}$ scales with $\nu_{\rm max}$, they all followed the practice of setting $W \simeq 2\Gamma_{\rm env}$, so that the full range encompasses most (strictly, just over 95\,\%) of the total oscillation power $P_{\rm tot}$. We proceed below to show that this is a sub-optimal choice.

\section{Optimisation of the full range $W$}
\label{sec:range}

\subsection{Generalised formulae for different ranges $W$}
\label{sec:forms}

To test how the global SNR and detection probability $p_{\rm final}$ depend on the choice of range $W$, we begin by presenting formulae that capture differing fractions of the full Gaussian envelope. We define a coefficient $\alpha$ that controls the range according to
 \begin{equation}
 W = \alpha \Gamma_{\rm env}.
 \label{eq:alpha}
 \end{equation}
Previous practice has therefore been consistent with setting $\alpha  \simeq 2$. 

The integrated area of the Gaussian envelope contained within a symmetric frequency range $W$ about $\nu_{\rm max}$ is 
 \begin{equation}
 P_{\rm tot}(W) = \int \limits_{\nu_{\max}-W/2}^{\nu_{\max}+W/2} P(\nu)\,\mathrm{d}\nu.
 \end{equation}
Evaluating the integral gives
\begin{equation}
 P_{\rm tot}(W) = P_{\rm tot} \operatorname{erf} \left(\frac{W \sqrt{\ln 2}}{\Gamma_{\rm env}} \right),
 \end{equation}
where $P_{\rm tot}$ is the full Gaussian area given by \eqref{eq:ptot}. We may rewrite the error function (erf) in the above in terms of $\alpha$ alone (compare \eqref{eq:alpha}), and changing the independent variable specifying the dependence of the integral on the selected range from $W$ to $\alpha$, we have
 \begin{equation}
 P_{\rm tot}(\alpha) = P_{\rm tot} \operatorname{erf} \left( \alpha \sqrt{\ln 2} \right).
 \end{equation}
The global SNR is
 \begin{equation}
 {\rm SNR_{\rm tot}} (\alpha) = \frac{P_{\rm tot}(\alpha)} {B_{\rm tot}(\alpha)},
 \end{equation}
which we may write explicitly as
 \begin{equation}
 {\rm SNR_{\rm tot}} (\alpha) = \frac{P_{\rm tot}}{\alpha\Gamma_{\rm env}\bar{b}(\alpha)}  \operatorname{erf} \left( \alpha \sqrt{\ln 2} \right).
 \end{equation}

To help understand the results that follow, it is also instructive to express the formulae in terms of an average power spectral density $\bar{P} (\alpha)$ across the range defined by $\alpha$, that is,
 \begin{equation}
 \bar{P} (\alpha) =  P_{\rm tot}(\alpha) / W \equiv \left( \frac{P_{\rm tot}}{\alpha \Gamma_{\rm env}} \right) \operatorname{erf} \left( \alpha \sqrt{\ln 2} \right).
 \end{equation}
The global SNR may then be written in the form
 \begin{equation}
 {\rm SNR_{\rm tot}} (\alpha) = \frac{\bar{P} (\alpha) }{\bar{b}(\alpha)},
 \end{equation}
from which it is clear that it scales directly with the average power spectral density. 

It is obvious that by increasing $\alpha$, and hence $W$, the average power spectral density $\bar{P} (\alpha)$ and the global SNR will decrease. However, this must be compared against the commensurate increase in $N$, and hence, the number of degrees of freedom governing the statistics of the global SNR. At fixed SNR, the larger $N$, the higher the detection probability $p_{\rm final}$. In short, the SNR required for a detection will be reduced. The combination of these opposing effects suggests that there will be an optimal choice for $\alpha$ that provides the best compromise.

\subsection{The optimal frequency range}
\label{sec:opt}

\begin{figure*}
\centering
\centerline{\includegraphics[scale=0.6]{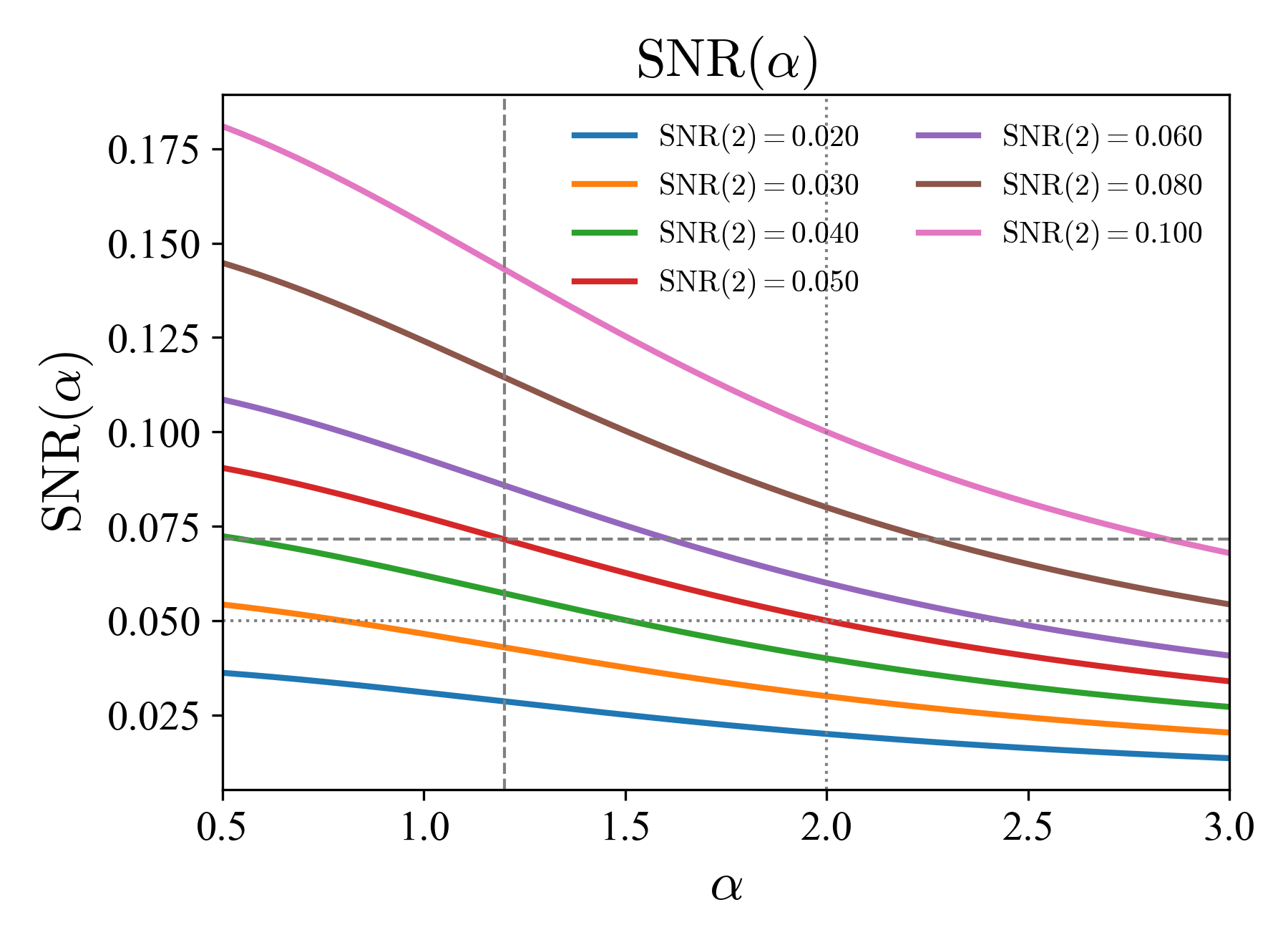}
            \includegraphics[scale=0.6]{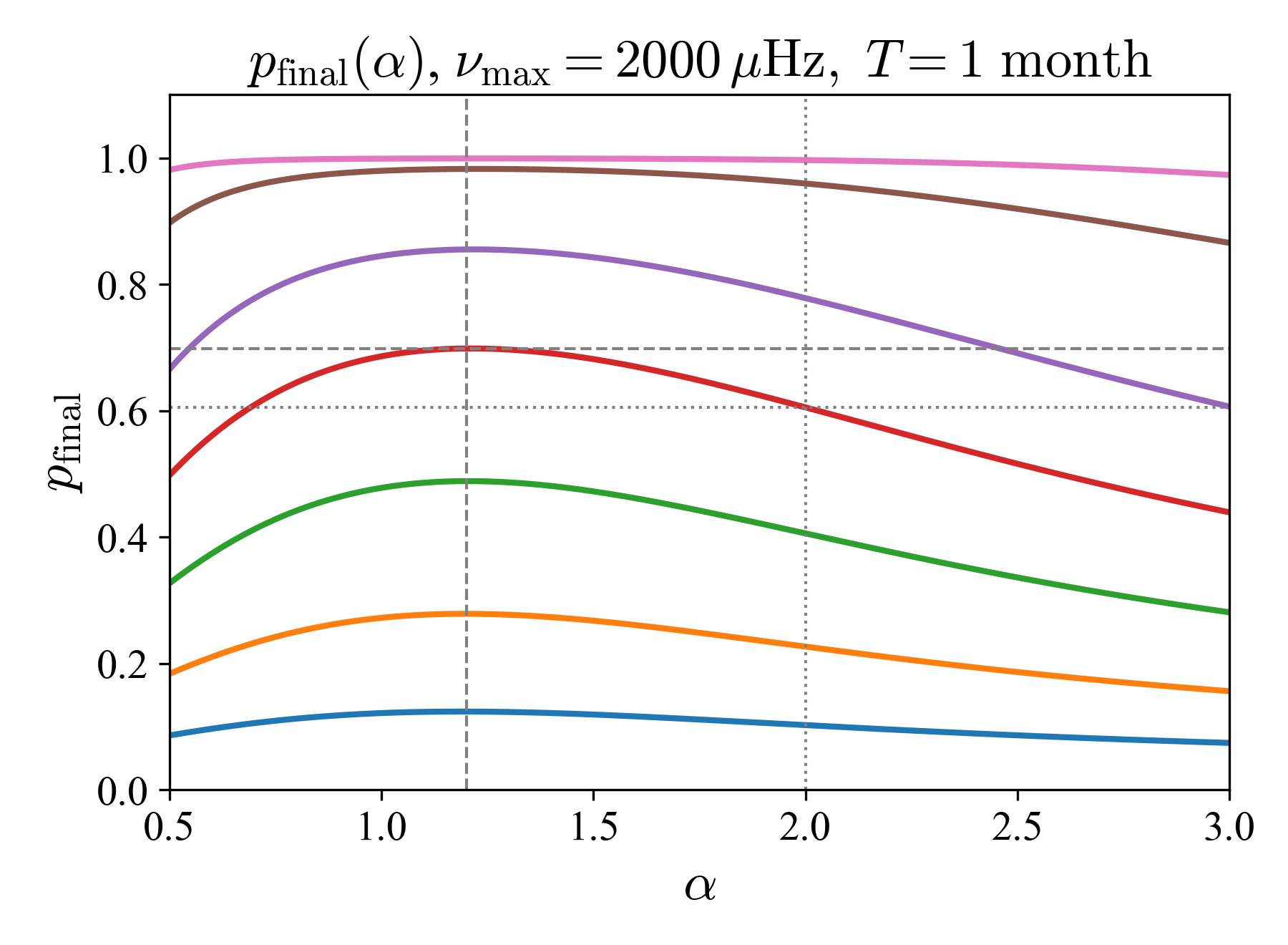}}
      \caption{Left: Variation in SNR values with $\alpha$ for different assumed values of $\rm SNR_{tot}(2) \in[0.02,0.10]$ (see plot annotation), with the vertical dotted line marking the standard value of $\alpha=2$ and the vertical dashed line marking the optimal value of $\alpha\simeq1.2$. The corresponding horizontal dotted and dashed lines mark the intersection with the red curve, having $\rm SNR_{tot}(2)=0.050$. Right: Corresponding detection probabilities, $p_{\rm final}$, for $\nu_{\rm max} = 2000\,\rm \mu Hz$ and $T=1$\,month.}
      \label{fig:plot1}
\end{figure*}

The left panel of \fref{fig:plot1} shows the variation in the global SNR with $\alpha$ for different assumed values of $\rm SNR_{\rm tot}(2) \in[0.02,0.10]$ (see plot annotation), with the vertical dotted line marking the standard value of $\alpha=2$. The plotted lines show the expected trend: $\rm SNR_{\rm tot}(\alpha)$ decreases with increasing $\alpha$. The right panel plots the corresponding detection probabilities, $p_{\rm final}$, assuming a false-alarm threshold of $p_{\rm false} = 0.01$. We used $\nu_{\rm max} = 2000\,\rm \mu Hz$ and calculated the envelope FWHM using the well-established relation $\Gamma_{\rm env} = 0.66 \nu_{\rm max}^{0.88}$ \citep{Mosser2012}. We also assumed observations of duration $T=1$\,month, from which we calculated $N$ for each value of $\alpha$ using Eqs.~\ref{eq:N} and~\ref{eq:alpha}. 

The right panel reveals a maximum in the curves at a value of $\alpha \simeq 1.2$, marked with the dashed line. The standard value of $\alpha=2$ is seen to be sub-optimal. Other choices of $\nu_{\rm max}$ and $T$ give essentially the same results, with some very small scatter in the optimal value for $\alpha$. The largest gain appears to come at intermediate detection probabilities. 


\begin{figure*}
\centering
\centerline{\includegraphics[scale=0.6]{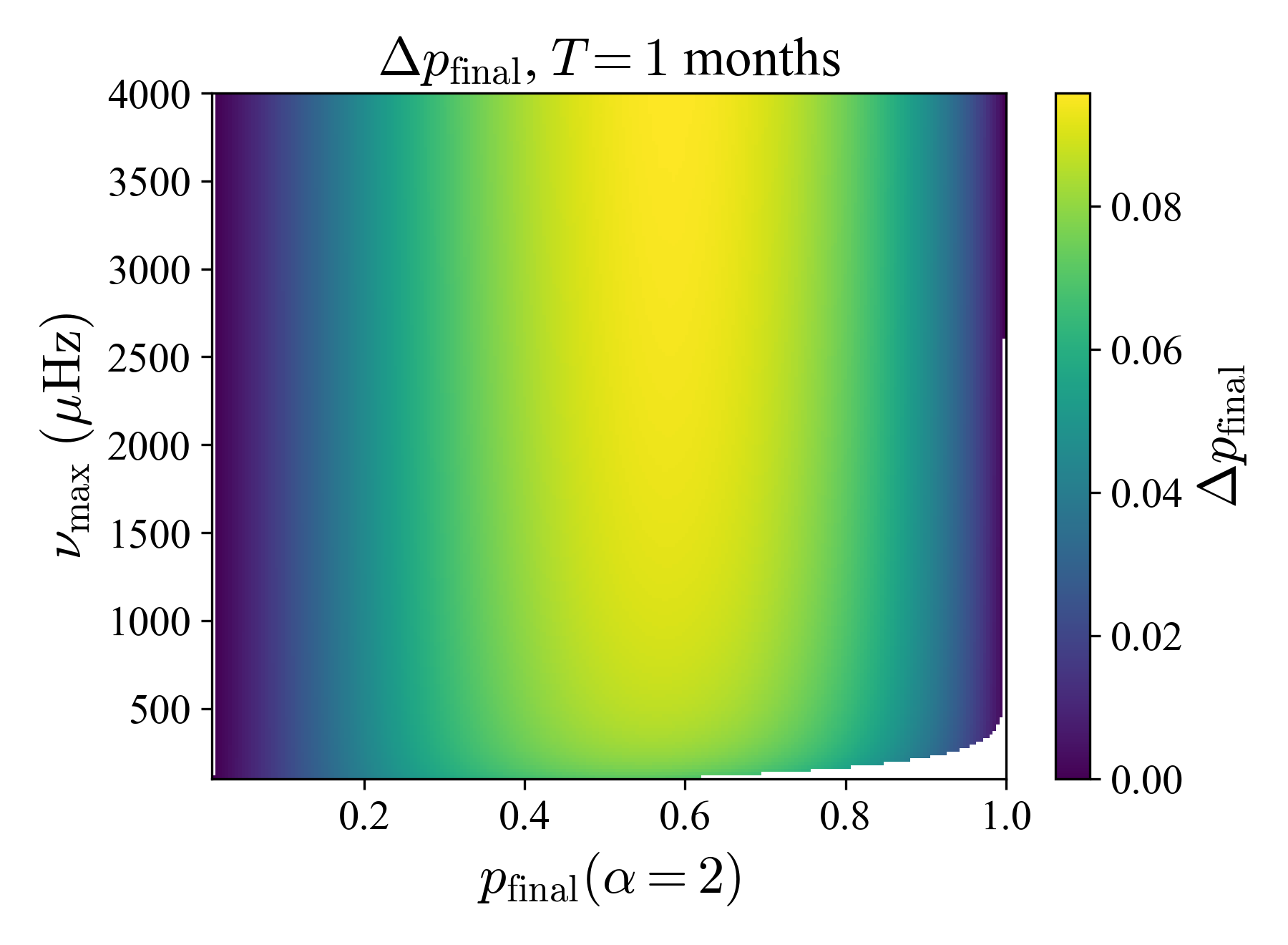}
            \includegraphics[scale=0.6]{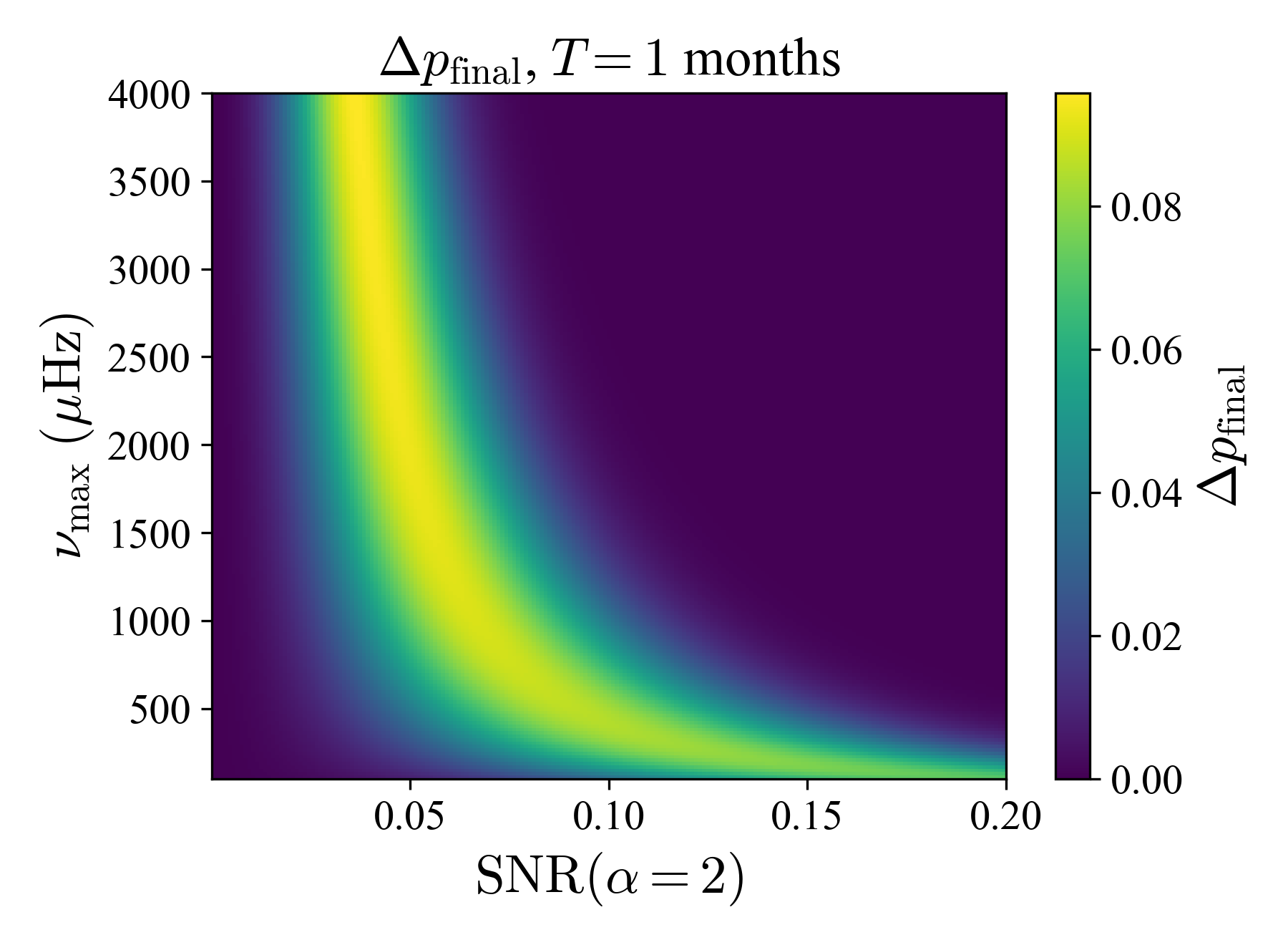}}
\centerline{\includegraphics[scale=0.6]{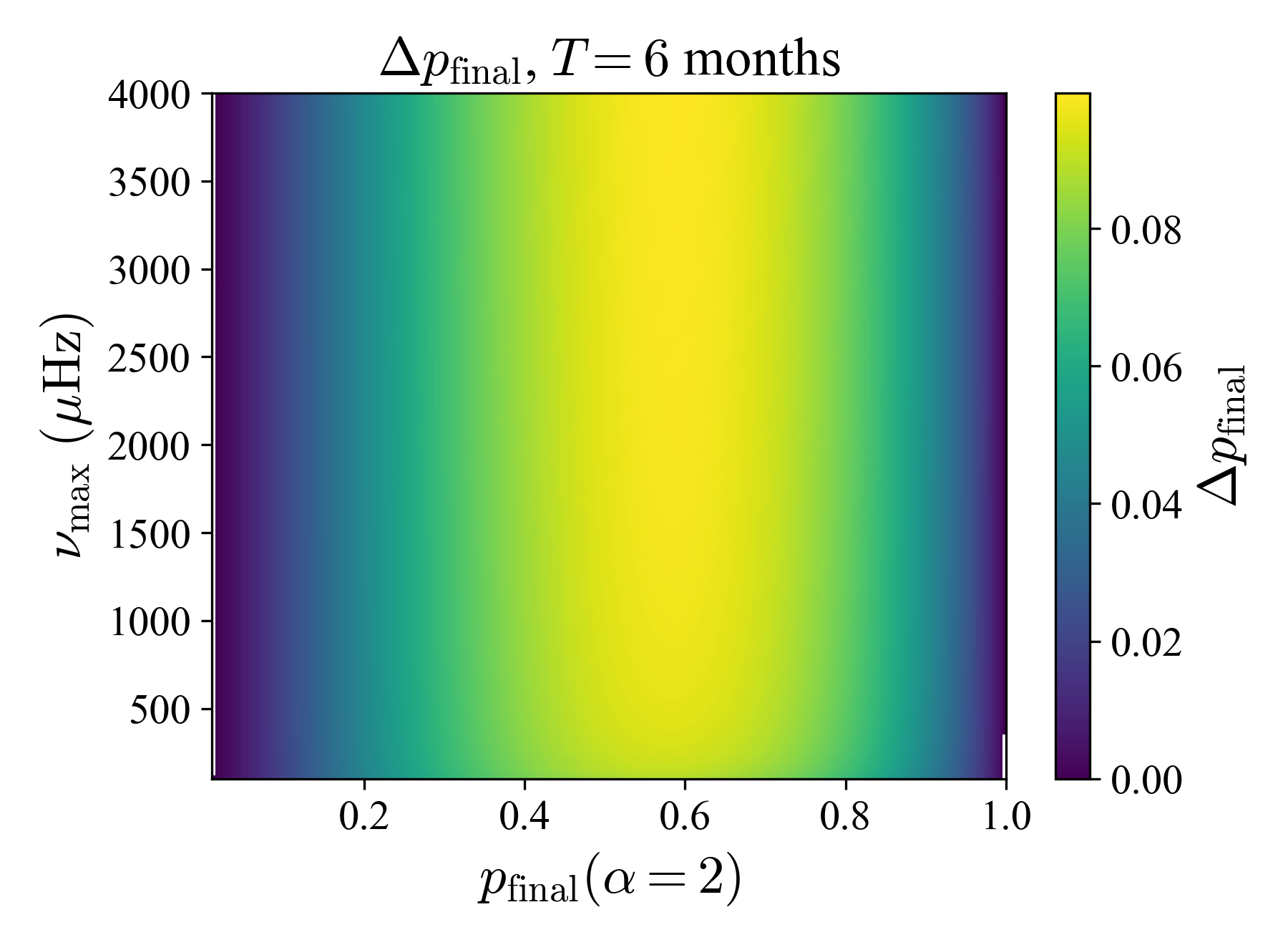}
            \includegraphics[scale=0.6]{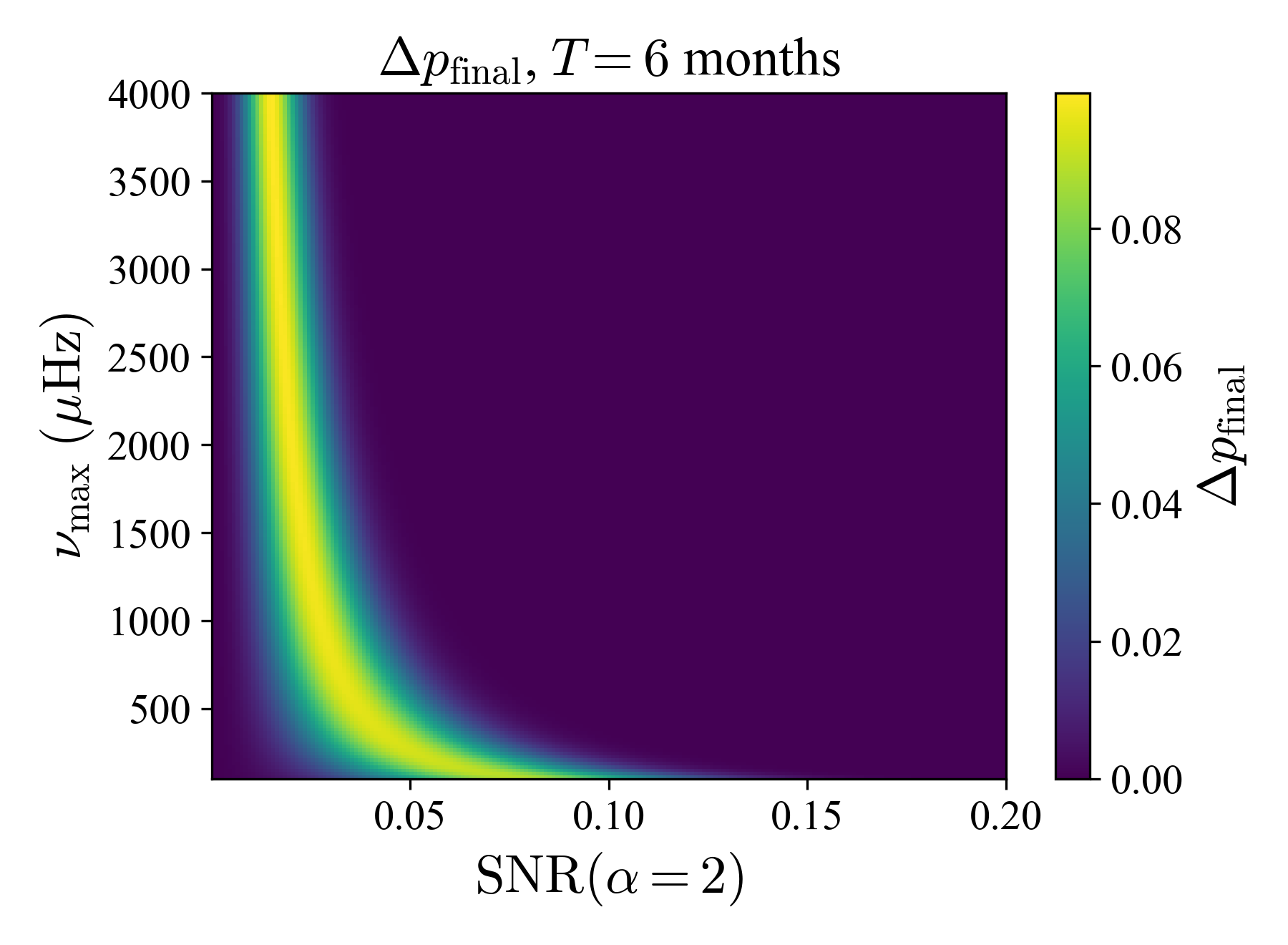}}
      \caption{Variation in the change $\Delta p_{\rm final}$ between $\alpha=1.2$ and 2.0 with $\nu_{\rm max}$, $p_{\rm final}$ (left column) and $\rm SNR_{\rm tot}(2)$ (right column). The top row shows results for $T=1$\,month, and the bottom row shows them for $T=6$\,months.}
      \label{fig:plot2}
\end{figure*}

This behaviour is illustrated in \fref{fig:plot2}, which shows how the change $\Delta p_{\rm final}$ between $\alpha=1.2$ and 2.0 varies with $\nu_{\rm max}$, $p_{\rm final}$ (left column) and $\rm SNR_{\rm tot}(2)$ (right column). The top row shows results for $T=1$\,month, and the bottom row shows them for $T=6$\,months. The right panel reveals a well-constrained optimal region in $\nu_{\rm max}$-SNR space. To the right of this region, the detection probability is already high, meaning little if any gain is to be had through changing $\alpha$. The upper right half of the plot will, in practice, not contain real stars, because the maximum mode amplitudes and hence $P_{\rm tot}$ decrease strongly with increasing $\nu_{\rm max}$. To the left of the optimal region, the SNR is so low that there is a negligible impact on the probability from changing $\alpha$. The optimal region moves to higher SNR at lower $\nu_{\rm max}$ because $N \propto \Gamma_{\rm env} \propto \nu_{\rm max}^{0.88}$ drops, meaning a higher SNR is required to give similar gains in probability. It is also shifted to higher SNR at lower $T$ for the same reason.
We note that increasing the false-alarm probability $p_{\rm false}$ has the effect of moving the ridge in $\Delta p_{\rm final}$ from \fref{fig:plot2} to lower $\rm SNR_{\rm tot}(2)$ values, and vice versa for a decreased $p_{\rm false}$.


\begin{figure*}
\centering
\centerline{\includegraphics[scale=0.6]{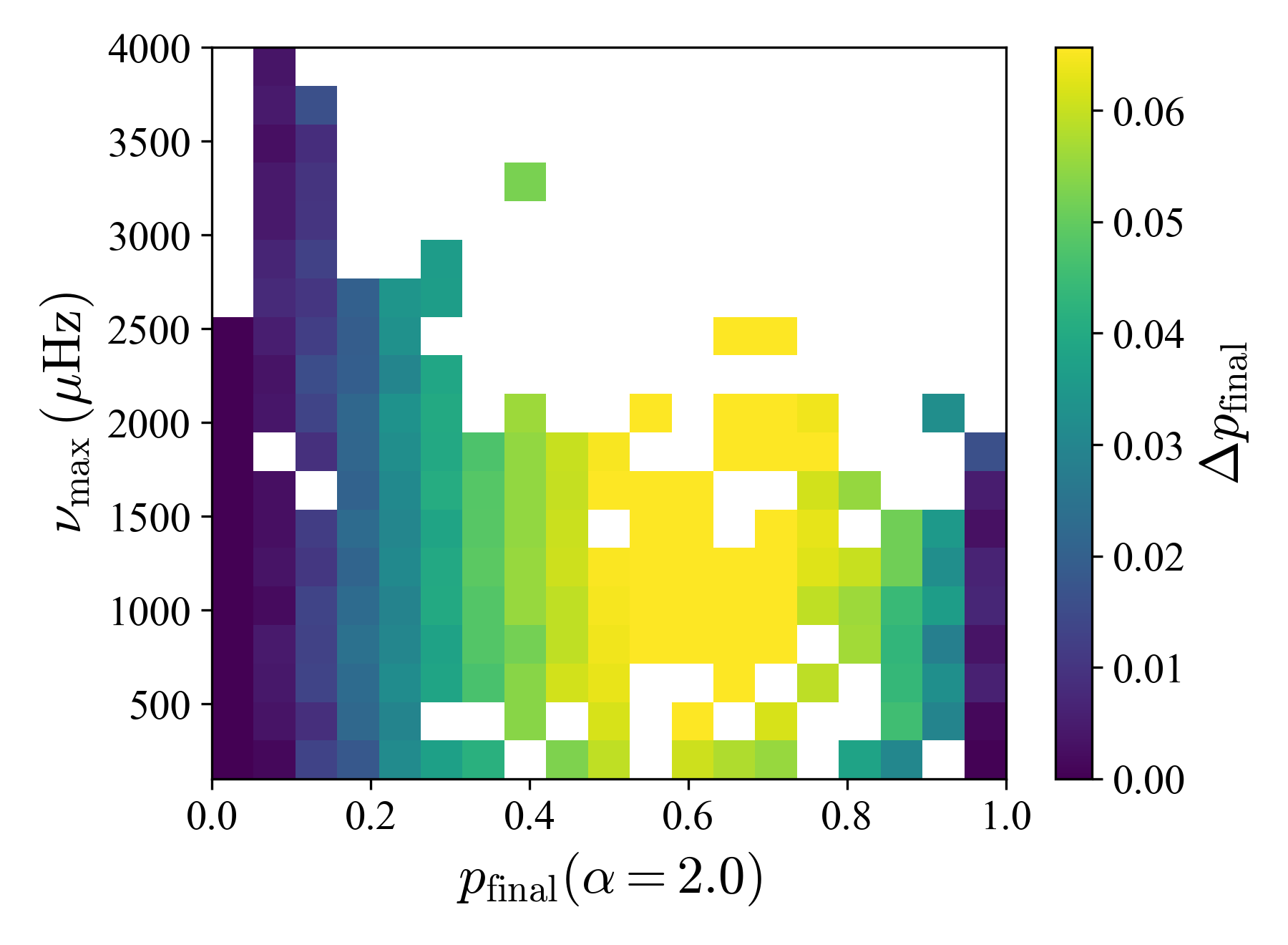}
            \includegraphics[scale=0.6]{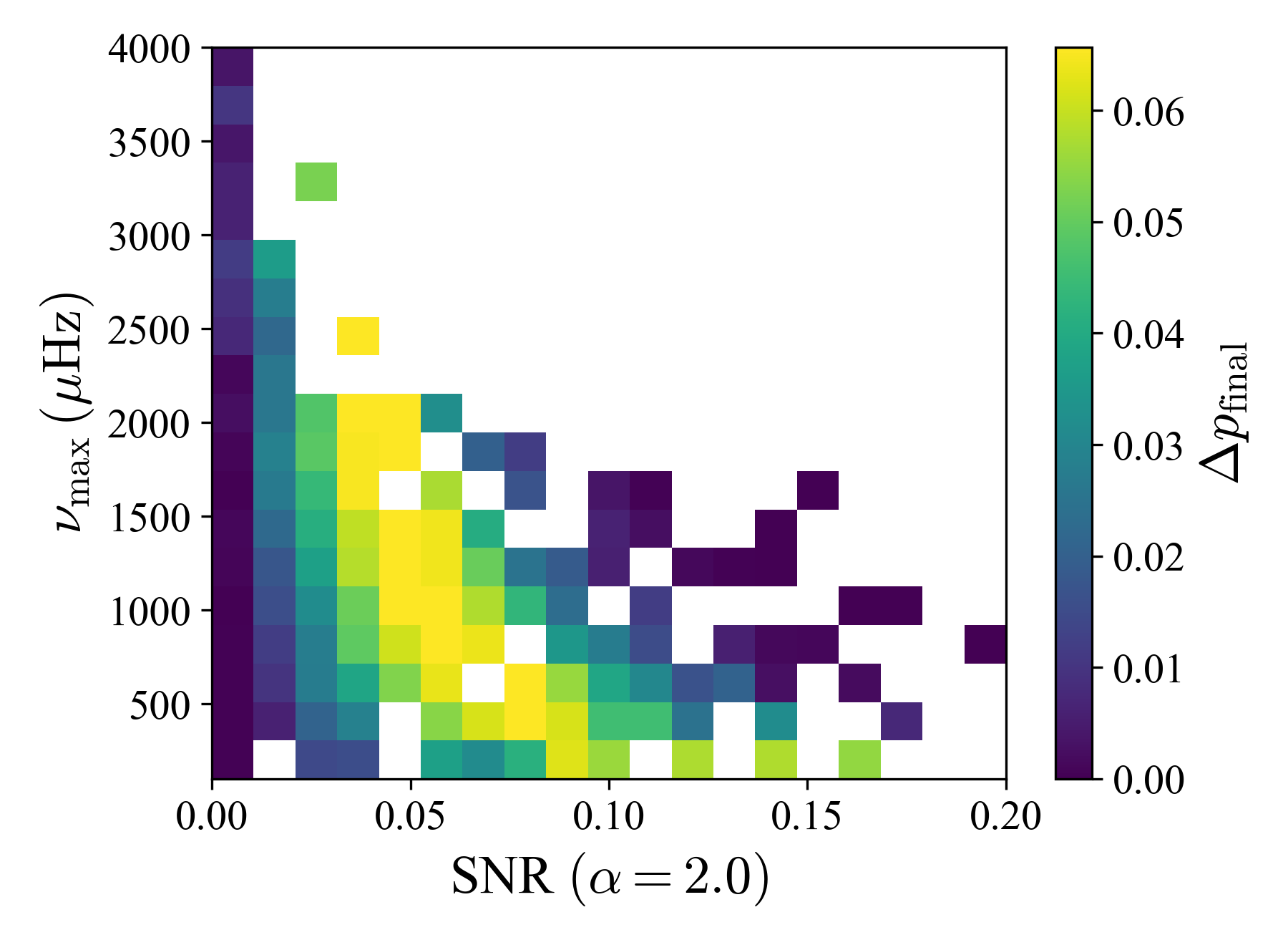}}
      \caption{Results for predictions using the bright TESS stars from the TESS Luminaries Sample \citep{Lund2025}, assuming a single sector of TESS observations ($T \simeq 1$\,month), data with a cadence of $120$ sec, a false-alarm probability of $5\%$, and input values of \teff, \logg, and $R$ from the Gaia DR3 \texttt{gspphot}.}
      \label{fig:plot3}
\end{figure*}

\Fref{fig:plot3} shows results of detectability predictions for a sample of bright stars observed by TESS \citep{Ricker2015}. Specifically, we adopted the $V\leq6$ stars from the TESS Luminaries Sample \citep{Lund2025}, which also includes the sample of bright evolved stars (their Fig.~1). Detectability predictions were made using the \texttt{ATL3} tool of \citet{Hey2024}\footnote{Updated with a TESS-specific bolometric correction for mode amplitudes \citep{Lund2019}.}, which implements the \citet{Chaplin2011} method, assuming a single sector of TESS observations ($T \simeq 1$\,month), data with a cadence of $120$ sec, a false-alarm probability of $5\%$, and input values of \teff, \logg, and $R$ from the Gaia DR3 \texttt{gspphot} \citep{Gaia2023}. \Fref{fig:plot3} shows the same structure as in the above calculations and plots, as well as the expected absence of stars in the upper right half of the $\nu_{\rm max}$-SNR plot.    

For this specific sample, the predicted yield for $p_{\rm final} \ge 0.90$ increased by ${\sim}12\%$ when we changed from $\alpha=2$ to $1.2$, and by ${\sim}5\%$ for $p_{\rm final} \ge 0.99$. This illustrates the benefits that accrue when considering the returns from a target ensemble. The improvement in the predicted yield strongly depends on the sample. For example, when a significantly longer observing duration is considered, many stars in a bright sample like the TESS one considered here would move to higher values of $\rm SNR_{\rm tot}(2)$ (to the right in \fref{fig:plot2}), and the relative improvement in $p_{\rm final}$ from changing $\alpha$ would decrease. Similarly, for a sample of dimmer stars, many would have very low $\rm SNR_{\rm tot}(2)$ values, and a longer observing duration $T$ would be needed to see a significant relative change in $p_{\rm final}$ from changing $\alpha$.


\section{Conclusion}
\label{sec:conc}

We have presented a straightforward optimisation of the asteroseismic detection prediction recipe of \citet{Chaplin2011} by tuning the frequency range used for assessing the significance of excess power from solar-like oscillations. Instead of the range normally adopted, corresponding to twice the FWHM of the oscillation envelope (\ie $2\Gamma_{\rm env}$), we found that adopting a range of $1.2\Gamma_{\rm env}$ maximises the probability of detection.

Choices made over which scaling relations to adopt in the calculation of the mode and granulation parameters can clearly change the calculated probabilities, both positively and negatively. However, changing the adopted frequency range from $2\Gamma_{\rm env}$ to $1.2\Gamma_{\rm env}$ moves the predictions in one direction only, that is, to higher probabilities. Depending on the sample considered, this optimisation can have a substantial effect on the predicted detection yield. We therefore recommend the modification be implemented in target selection and detection prediction efforts for future missions suitable for asteroseismology, including the ESA PLATO Mission \citep{Goupil2024}, NASA's Nancy Grace Roman Space Telescope \citep{Weiss2025}, and the proposed ESA High-precision AsteroseismologY of DeNse stellar fields (HAYDN) mission \citep{Miglio2021}. Furthermore, it is important to add that using our recommended range will also be beneficial for codes that analyse the data with the aim to detect solar-like oscillations, based on testing the significance of excess mode power \citep[\eg][]{Appourchaux2004,Lund2012,Bell2019,Viani2019,Nielsen2022}, since its adoption will also optimise the probability of making robust detections. 


\begin{acknowledgements}
MNL acknowledges support from the ESA PRODEX programme (PEA 4000142995). WJC acknowledges the support of the UK Space Agency. This research has made use of NASA's Astrophysics Data System Bibliographic Services. We thank the referee for their helpful report.

\end{acknowledgements}

\bibliographystyle{aa} 
\bibliography{aadet1.bib} 

\end{document}